\documentclass[preprint,prb,showpacs]{revtex4}
\usepackage{graphics}
\usepackage{graphicx}
\usepackage[pdftitle={Heusler Compounds 2006},
           pdfauthor={Fabio Ricci}
           ]{hyperref}
\usepackage[dvips]{color}
\usepackage{color}
\usepackage{fancyhdr}
\usepackage{epsfig}
\newcommand{\cmg}{Co$_2$MnGe}

\newcommand{\cms}{Co$_2$MnSi$\:$}
\newcommand{\cmsn}{Co$_2$MnSn$\:$}
\begin{document}

\title{Optical and magneto-optical properties of ferromagnetic
full-Heusler films: experiments and first-principles calculations}
\author{F. Ricci$^{1}$, S. Picozzi$^{2,1}$, A. Continenza$^{1,3}$, F. D'Orazio$^{1,3}$, F. Lucari$^{1,2}$, \\ K. Westerholt$^{4}$,
M. Kim$^{5,}$\footnote{Present address:
Seoul National University, Seoul,
Korea} and A. J. Freeman$^{5}$ }
\affiliation{$^{1}$ Dip. di Fisica, Univ. dell'Aquila, 67010 Coppito (L'Aquila),
Italy;\\
 $^{2}$ Istituto Nazionale di Fisica della Materia (INFM-CNR);\\
$^{3}$ Consorzio Nazionale Interuniversitario Scienze della Materia (CNISM);\\
$^{4}$ Institut fur Experimentalphysik IV, Ruhr-Univ. Bochum, Germany;\\
$^{5}$ Dept.$ \ $of Physics and Astronomy, Northwestern Univ., 60208 Evanston, Il, USA}

\begin{abstract}
We report a joint theoretical and experimental study focused on understanding the optical and magneto-optical
properties of Co-based full-Heusler compounds. We show that
magneto-optical spectra calculated within \emph{ab-initio} density functional theory are able to uniquely identify the features of the
experimental spectra
in terms of spin resolved electronic transitions.
As expected for 3$d$-based magnets, we find that the largest Kerr rotation for these alloys is of the order of 0.3$^{o}$
in polar geometry. In addition, we demonstrate that (\emph{i}) multilayered structures have to be carefully handled in the theoretical calculations
in order to improve the agreement with experiments, and (\emph{ii})
combined theoretical and experimental investigations constitute a powerful approach to designing new materials for magneto-optical and spin-related
applications. 

\end{abstract}
\maketitle
\section{Introduction}

Spin electronics, or spintronics, is a multidisciplinary field that involves the active manipulation of
the electron spin degree of freedom. Recent developments in material design allow to combine magnetic and
semiconducting character, thus opening the possibility to build devices with highly polarized
currents. In fact, one of the major challenges in spintronics is
the search for compounds
that could act as efficient injectors of spin-polarized current
in semiconductors. Diluted magnetic semiconductors (DMS)\cite{dietl} 
are obviously
among the best candidates as spin-polarized sources, 
due to their excellent structural
compatibility with the spin-current semiconducting 
``sink". Indeed, spin-light emitting devices have been successfully 
realized using DMS as spin-injectors\cite{molenk}.
However, the very low Curie
temperatures
and the need for high magnetic fields 
seem  to have  hindered so far
a large-scale application of DMS in the spin-injection
framework.
On the other hand, the discovery of a wide class of half-metallic (HM)
compounds, initially predicted by de Groot \emph{et al.}\cite{degroot}, has gained importance in this
context\cite{galana}. The Co$_2$MnX (with X = Al, Si, Ga, Ge, Sn) full-Heusler family has interested the scientific
community because of its ideal
100\% spin polarization at the Fermi level, a good structural matching 
with mainstream semiconductors (in terms of lattice constants
and crystal structure)
and very high Curie temperatures\cite{galanak,picozzi} ($\sim$ 900 K).

Recently, a number of Heusler compounds have been grown epitaxially on
semiconducting substrates: for example, \cmg\cite{ambrose}, \cms\cite{raphael,prinz}, 
Ni$_2$MnGa\cite{nimnga}, Ni$_2$MnGe\cite{nimnge} on
GaAs [001] and Ni$_2$MnIn on InAs [001]\cite{nimnin}.
On the other hand, also RF-sputtering technique has been shown to produce homogeneous Heusler films with good
structural and magnetic quality\cite{geier,geier2}.  
The first noticeable application of electrical spin injection
from Co$_2$MnGe into a $p-i-n$ Al$_{0.1}$Ga$_{0.9}$As/GaAs light emitting diode heterostructure
was reported only recently\cite{palmstrom}: a steady state polarization of 13$\%$ resulted at 2 K.
The injected polarization was significantly lower than
the theoretically predicted 100$\%$ value expected for half-metals 
\cite{ishida,picozzi}, pointing to
the importance of interfacial  disorder or temperature effects\cite{ishida2}. 

Moreover, within this family of compounds, PtMnSb has been found to exhibit a giant
Kerr rotation (up to 2.0$^{o}$ at room temperature and 
5.0$^{o}$ at 80 K) that is totally unexpected
for a 3$d$-based material\cite{ptmnsb}. This is of technologically relevant
importance in the context
of magneto-optical reading and recording technology.
Since then, there has been great interest in magneto-optics for this 
class of compounds\cite{revantonov}; in particular, \cms\cite{wang} and \cmg\cite{xmcdge,xmcdge2,xmcdge3} were characterized by means of
X-ray absorption and magnetic circular dichroism. 
However, to our knowledge,
there are no experimental or first-principles characterizations (within the local-spin
density approximation (LSDA) to density functional theory (DFT)) 
of these
Co-based compounds, as far as the magneto-optical Kerr effect (MOKE) is concerned. 

This paper presents a joint theoretical and experimental study of Heusler compounds, aimed at a more complete
understanding of their optical and magneto-optical properties. Therefore, we pick two prototype examples of the full-Heusler family
(Co$_2$MnGe and Co$_2$MnSn), which are similar enough to support the comparison between calculated
and experimental MOKE spectra. These crystals are expected to have a high degree
of spin-polarization at the Fermi level ($\sim$100\%), and a similar response when they are excited with electromagnetic radiation.
 
The paper is organized as follows: after some computational details are reported in Sec. \ref{techn}, we focus  
  on the description of the experimental approach to the magneto--optical
   properties (Sec. \ref{expt}) and
  on the comparison between theory
and experiments for the
Kerr rotation and ellipticity 
(Sec. \ref{theoexpt}). 
The good agreement with the experimental spectra suggests that a first principles
calculation describes sufficiently
well the Heusler's electronic structure 
on the basis of a bare DFT approach, without invoking correlation or excited-state
 effects beyond a standard LSDA treatment. Therefore, in Sec. \ref{theory}, we
 give a comprehensive theoretical description of both optical and
 magneto--optical properties for Co$_2$MnGe and \cmsn on the basis of their underlying
 electronic structure.  In Sec. \ref{concl} we draw conclusions.

\section{Computational details}
\label{techn}
Accurate
first-principles calculations, using the full-potential linearized augmented plane-wave (FLAPW) method\cite{flapw},
  were performed within the LSDA approximation 
to DFT, according to the von Barth-Hedin treatment for the exchange--correlation functional.
In the calculations, we used 
muffin tin radii of 2.1 a.u. for Ge, Co, Mn and 2.3
a.u. for Sn; wave function and charge density cut-offs were set at 3.8 and 8.0 a.u., respectively
and 110 {\bf k}-points were used in the irreducible
wedge of the Brillouin zone for the self-consistent cycle, according to 
the standard Monkhorst-Pack procedure\cite{monkh}.
 
We recall that MOKE is based on the change experienced by linearly polarized light reflected from a 
ferromagnetic material, which 
becomes elliptically polarized with its major axis rotated by an angle $\theta_{\kappa}$ (with respect to the 
incident polarization plane) and ellipticity $\varepsilon_{\kappa}$\cite{buschow}. 
The MOKE interband transition is a direct consequence of the interplay
between spin-orbit coupling and the exchange interaction: the relation between the magnetic field,
acting on spins, and radiation, which interacts with the orbital degrees of freedom,
is established through spin-orbit coupling. The symmetry breaking resulting from this
interaction changes the electronic selection rules of the system in such a way that most
of the important features in the conductivity tensor $\sigma(\omega)$ can be interpreted in terms of the
bare band structure (see Sec. \ref{theory}). 


The optical conductivity 
tensor $\sigma$
is calculated within linear response theory in the 
long wavelength limit, using the relativistic single-particle Kohn-Sham eigenvalues and wave 
functions\cite{callaway}. Linear response theory within the Kubo formalism allows us to write
the conductivity tensor $\sigma$ as:
\begin{equation}\label{eq:sigma}
\sigma_{\lambda\lambda'}(\omega)=\frac{ie}{m^{2}\hbar V}
\sum_{(i\vec{k})_{occ}}^{(j\vec{k})_{nocc}}\frac{1}{\omega_{ij}}\left[\frac{\Pi^{\lambda}_{ij}\Pi^{\lambda'}_{ji}}{\omega-\omega_{ij}+i/\tau}-\frac{\Pi^{\lambda}_{ij}\Pi^{\lambda'}_{ji}}{\omega+\omega_{ij}+i/\tau}\right]\quad,
\end{equation}
where the sum is taken between the occupied and unoccupied states, $i,j$ respectively; $\Pi^{\lambda}_{ij}$
is the $\lambda$ component ($\lambda=x,y,z$) of the dipole matrix calculated between the $i,j$ states;
$\tau$ is the interband relaxation time.
In polar geometry at normal incidence ({\em i.e.} with
the propagation direction ($z$ axis) parallel to the magnetization direction and perpendicular to the
surface), for cubic crystals,
the complex polar Kerr angle can be expressed to first order as:
\begin{equation}
\label{eq:teta3}
\Phi_{\kappa} = \theta_{\kappa} + i\varepsilon_{\kappa} =
\frac{-\sigma_{xy}}{\sigma_{xx}\sqrt{1 + i(4 \pi \sigma_{xx}/\omega)}}=
\frac{-\sigma_{xy}}{D}
\label{boh}
\end{equation}
where $\omega$ is
the photon energy and we define the quantity $D$ containing only the diagonal terms
of the optical conductivity.
We account for other transitions that are not included in the {\sl interband} term expressed in Eq. (\ref{eq:sigma})
(i.e. scattering at lattice imperfections or phonons)\cite{ebert} by considering {\sl intraband}
transitions described by an empirical Drude contribution (characterized by an intraband
relaxation time $\tau_D$) added to the diagonal elements of the $\sigma$ tensor:
\begin{equation}
\label{eq:drude}
\sigma_{0}=\frac{\omega_p^2}{4\pi(1-i\omega\tau_D)}\quad .
\end{equation}
Here the plasma frequency is expressed as:
\begin{equation}
\omega_p=\frac{4\pi e^2}{m^2 V} \sum_{i{\bf k}, \lambda} \delta(\varepsilon_{i{\bf
k}}-E_F)|\Pi^{\lambda}_{ii}|^2\quad,
\end{equation}
where $E_F$ is  the Fermi energy, and $\Pi^{\lambda}_{ii}$ the diagonal element of the dipole matrix.

The calculation of the optical and magneto-optical properties has been performed
using 1000 {\bf k}-points.
In order to consider the effect of finite lifetimes, as well as of the experimental 
resolution, a Lorentzian broadening equal to $\frac{\hbar}{\tau}$ = 0.4 eV was applied for both
the interband and intraband contributions. Since MOKE is a very small effect, an accurate
all-electron code with no approximations on the potential, as the FLAPW used in the
present work, is highly recommended in order to obtain reliable results\cite{comkim}.


\section{Experimental results}
\label{expt}
Experimental MOKE spectra were obtained on \cmsn and Co$_2$MnGe films grown by RF-sputtering on Al$_2$O$_3$ a-plane substrates. Before deposition,
a 5 nm thick seed layer of  vanadium was deposited in order to induce excellent (110)-growth of the 100 nm thick Heusler
film\cite{geier}. A final protective overlayer of amorphous alumina, 5 nm thick, was then grown on top of the 
film.
At each photon energy complete hysteresis loops were performed and both $s$- and $p$-waves 
MOKE rotation and ellipticity were measured as a function of the magnetic field, in the range $\pm$5600 
Oe. In this way, the presence of paramagnetic or diamagnetic terms can be detected as a linearly field dependent 
component and then subtracted from the overall signal. This is particularly important for low temperature measurements 
in which the cryostat windows may provide an appreciable diamagnetic contribution.

As expected in thin films, due to shape anisotropy, magnetization in the direction perpendicular to the surface is not (energetically)
favored. This is evidenced by the linear field dependence of the polar MOKE hysteresis loops all over the magnetic
field range (not shown). On the contrary, longitudinal $s$-polarized MOKE experiments lead to hysteresis loops with a saturation field on
the order of 50 Oe.

\begin{figure}
\phantom{b}
\vspace{3cm}
\includegraphics[angle=-90,scale=0.65]{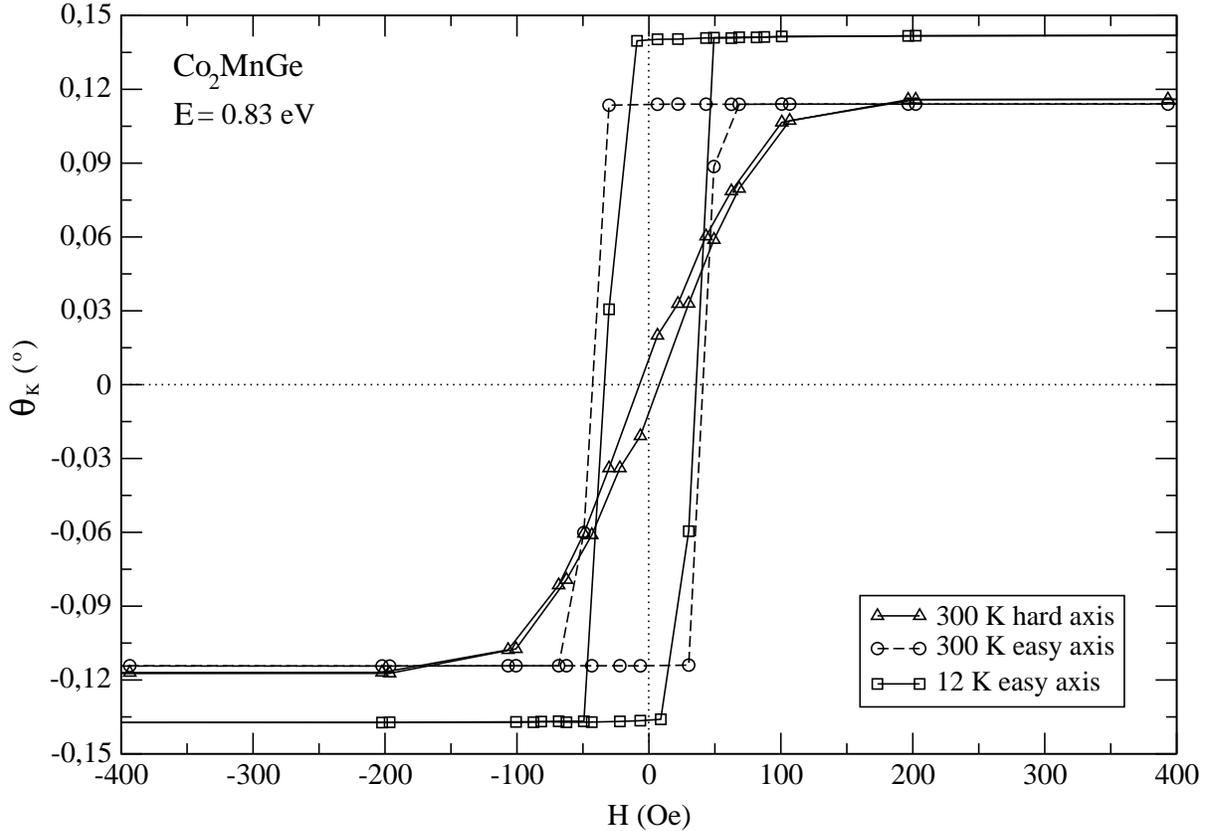}
\caption{$s$-polarized longitudinal MOKE rotation hysteresis loops of Co$_2$MnGe at photon energy = 0.83 eV.
Both easy and hard directions of magnetization are measured at room temperature.}
\label{fig:f2}
\end{figure}

As an example, we report in Fig. \ref{fig:f2} some $s$-wave longitudinal MOKE hysteresis loops for the Co$_2$MnGe$\ $ sample at a
photon energy of 0.83 eV. In-plane anisotropy is observed when longitudinal experiments 
are repeated with the sample rotated around the axis perpendicular to the surface; it turns out that the easy and hard axes are
mutually perpendicular. 
The in-plane anisotropy is induced during growth by the step edges on the surface of the Al$_2$O$_3$
substrate. However, since we 
are interested in comparing the experimental MOKE results with the theoretical prediction, the 
meaningful quantity to be considered is the saturation of the loops, which does not depend on the
in-plane angle and corresponds to the remanence measured
when the magnetic field lies along the easy axis direction.

\begin{figure}
\phantom{b}
\vspace{3cm}
\includegraphics[angle=0,scale=0.75]{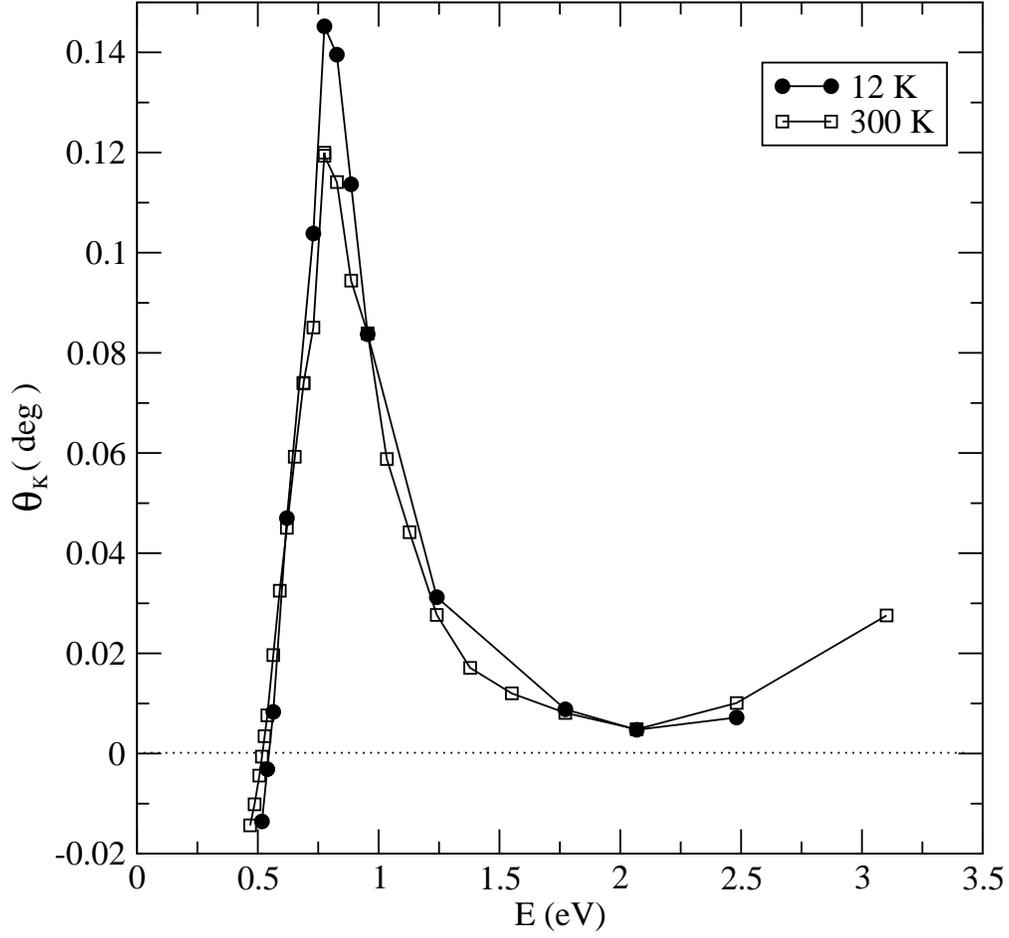}
\caption{$s$-polarized longitudinal Kerr rotation at saturation as a function of photon energy
at two different temperatures for \cmg.}
\label{fig:f3}
\end{figure}

The temperature 
dependence is shown more clearly in Fig. \ref{fig:f3}, where the complete spectrum of s-polarized longitudinal 
Kerr saturation rotation is reported at 300 K and 12 K for Co$_2$MnGe.
The main features are well 
reproduced at both temperatures,
except for a slightly higher peak observed at low temperature.
This is attributed to a reduced magnetization at high temperature. Similar 
outcomes are found for the \cmsn film as well. In the following, we examine in detail 
the results obtained at room temperature, since the presence of the 
cryostat windows and the undesirable vibrations of the sample holder lower the signal-to-noise ratio at low temperature,
thus preventing an accurate and complete determination of the spectra. It is also worthwhile  mentioning 
that room temperature magnetization measurements, performed with an alternating gradient 
magnetometer, provide a saturation magnetization corresponding to 83.06 emu/g and 74.38 emu/g 
for $\ $ and \cmsn, respectively. Not unexpectedly, these values are 20$\%$ and 15$\%$ lower, respectively, than 
the values obtained at 5 K on similar films reported previously\cite{geier}.

\section{Comparison between theory and experiments}
\label{theoexpt}

\begin{figure}
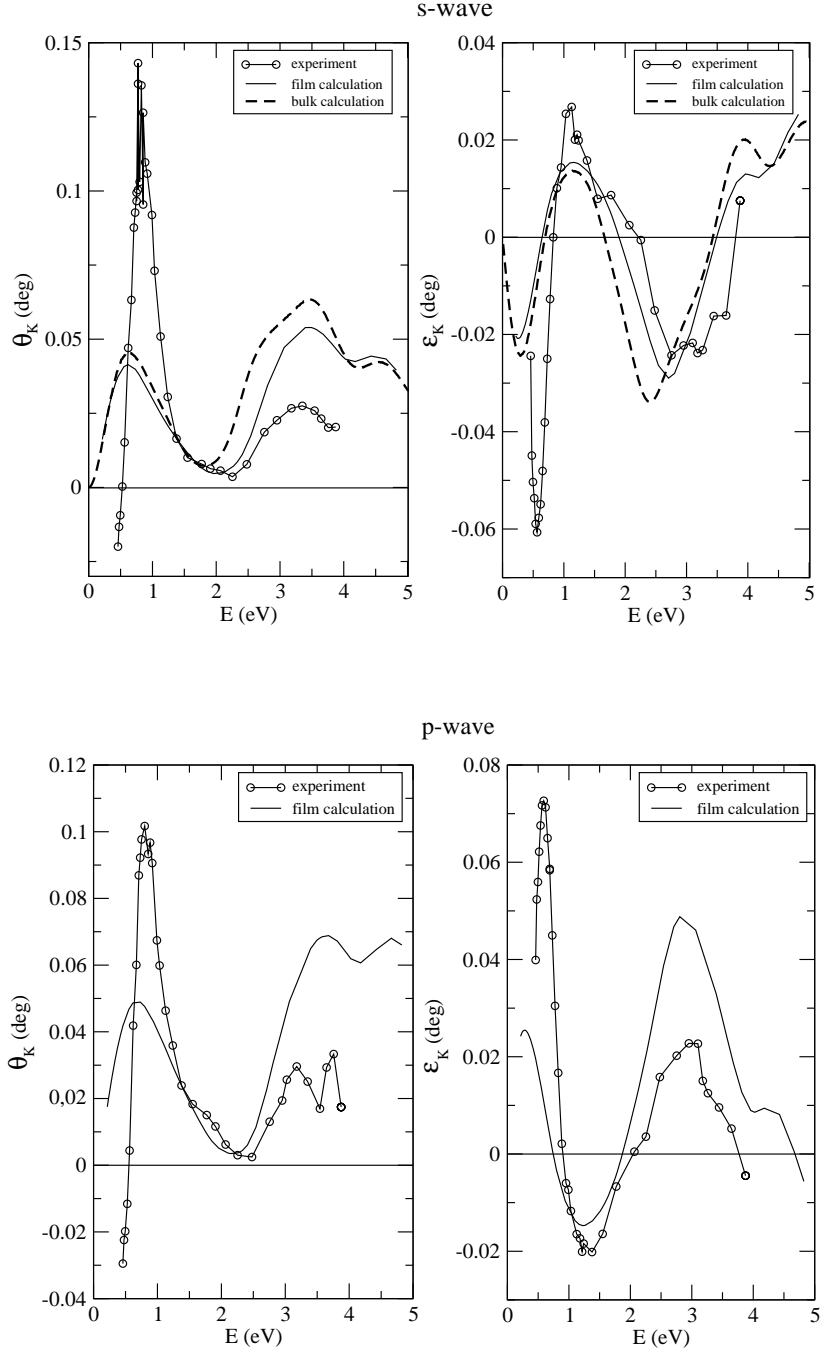

\includegraphics[angle=-0,scale=0.47]{Co2MnGe_s_film-bulk.eps}\\
\vspace{1.2cm}
\includegraphics[angle=-0,scale=0.47]{Co2MnGe_pwave.eps}
\caption{Theoretical and experimental longitudinal Kerr spectra as a function of photon energy in \cmg.
Top panels (s-wave) report both bulk (dashed)
and 100 nm thick film (solid) calculations.}
\label{fig:f4}
\end{figure}

In Figs. \ref{fig:f4} and \ref{fig:tre}, the complete experimental and theoretical spectra obtained for
longitudinal Kerr geometry are compared for the Co$_2$MnGe and \cmsn films, respectively. 
The different panels show the MOKE rotation and ellipticity, both in $s$ and $p$ polarization.
Our theoretical calculations are obtained according to the following procedure: the $\sigma$ conductivity tensor is calculated by
ab-initio simulations. Using Eq. (\ref{eq:teta3}), one can compute the Kerr angle at normal incidence
on the interface between two semi-infinite media: vacuum and \cmsn or Co$_2$MnGe.
However, in order to compare the experimental data with the theoretical predictions,
the calculated spectrum has to be evaluated in the same geometry and reflection conditions
(incident radiation at 45$^{o}$ on the sample surface) taking also into account the sample structural characteristics.
In particular, due to the finite 
thickness of the Heusler compound, we must consider that the reflected 
radiation is affected by the optical and magneto-optical response of a multilayered structure. 
This is possible according to the formalism introduced by Zak \emph{et al.}\cite{zak}: propagation
of the radiation within each layer is characterized by a propagation matrix P and a boundary matrix A,
which are functions of geometrical 
parameters, of the complex refractive index $n$ of the constituent material, and the complex magneto-optical
coefficient Q (if the layer is magnetic). The latter terms ($n$, Q) are related to the diagonal and off-diagonal 
elements, respectively, of the conductivity tensor calculated according to
Eq. (\ref{eq:sigma}). Using the P and A matrices, the boundary conditions
can be directly applied to the electric and magnetic field vectors that, depending on the light polarization ($s$ or $p$), satisfy
a different set of  homogenous non-linear equations. All results obtained within the Zak model are of the first order in Q. 
This framework takes into account multiple reflections at the interface
between sample layers and it allows us to predict the magneto-optical response of a ``film'' or a multilayered
structure. 
The wavelength 
dependence of the refractive index of the sapphire capping layer and substrate was taken from the literature\cite{index};
the vanadium seed layer was neglected since only a small contribution to the reflected radiation is expected from more internal thin layers.
For comparison, in the two upper panels of Fig. \ref{fig:f4}, the results calculated for the semi-infinite magnetic layer
(``bulk'' calculation) are included only for the case of $s$-wave in Co$_2$MnGe.
We observe that the experimental features (in particular the peak positions) are better matched by the film calculation (solid lines).

The intraband contribution to the conductivity tensor was modeled according to Eq. (\ref{eq:drude})
taking $\hbar\omega_p =$ 3.0 eV and $\frac{\hbar}{\tau_D} = $ 0.2 eV for both
samples, while for the interband term line-width, the value $\frac{\hbar}{\tau} = $ 0.4 eV was considered.
These values were chosen in order to fit the experimental spectra and they are all within the typical
range of values for these parameters; their energy dependence was neglected\cite{ebert}.
In particular, the above mentioned value for $\omega_p$
differs from the plasma frequency calculated
from first principles ({\em i.e.} $\hbar\omega_p \sim$ 4.4 eV for both compounds). 
However, this is not a serious discrepancy since the plasma frequency is known to be strongly affected by sample
 purity and it is absolutely
reasonable (if not expected) that disorder, neglected
in the $ab-initio$ simulations, exists in real samples and is responsible for the different $\omega_p$ value. 

Overall, for both compounds and for both $s$- and $p$-wave cases,
the agreement is good as far as the Kerr rotation and the ellipticity
are concerned: theory reproduces the exact sign of the angle and the spectral shape 
in the whole energy range examined. In particular,
the agreement between the energy positions of the main features is excellent, whereas the
magnitude shows a disagreement. This is really not too surprising if we consider that:
{\em i}) simulations are performed at zero
temperature and, therefore, they do not take into account spin fluctuations that occur
at room temperature as in the case of the actual measurements; {\em ii}) at low energies (E $\leq$ 1.5 eV ), intraband excitations
play an important role and may not be well reproduced by the simple Drude expression; 
{\em iii}) disorder in the Co and Mn sublattices occupation might also be present in real samples, thus affecting the electronic and magnetic
structure. For example, it has been theoretically shown that Co antisites
in bulk  \cmg $\ $ have low formation energy and destroy 
half-metallicity\cite{slvdefects}, at variance with the NiMnSb case where recent calculations showed that
low formation energy defects preserve 
the half metallic character\cite{alling}. However, our theoretical investigation allows us to obtain valuable insights
on the electronic properties of the materials, which are not easily accessible to experiment alone.

\section{Discussion of theoretical results}
\label{theory}


Before going into the details of the origin of the optical and
magneto-optical properties, we briefly recall
the structural and electronic properties 
of the different Heusler
compounds. In particular, in Table \ref{ricap} we report our
calculated\cite{ggavslda}
equilibrium lattice constant, 
along with the total magnetic moment, plasma frequency
and density of states at the Fermi level ($E_F$), the last two quantities
being spin-resolved. The plasma frequency of the majority-spin component is pretty large for both compounds, reflecting the 
highly dispersed character of the bands crossing $E_F$. 
We recall that, according to the previously calculated
electronic structure  for \cmsn  and \cmg\cite{slvdefects},
the majority spin DOS has metallic character, and the number of unoccupied electronic states
is low ($\sim$ 0.2 states/(eV cell)). On the other hand, the minority spin presents 
an occupied and unoccupied DOS (obviously separated by the 
gap giving rise to HM character), both with appreciable weight ($\sim$ 5 states/(eV cell)).
In addition, we observe that, in the energy range of interest
(from 0 to 5 eV), the dipole matrix elements for transitions 
within the minority channel are more than twice
as large than those in the majority spin channel, thus suggesting that optical
transitions will more likely occur in the minority spin channel. 
In the following analysis, therefore, we will mainly consider possible transitions
in the minority spin-channel (see below).

\begin{figure}
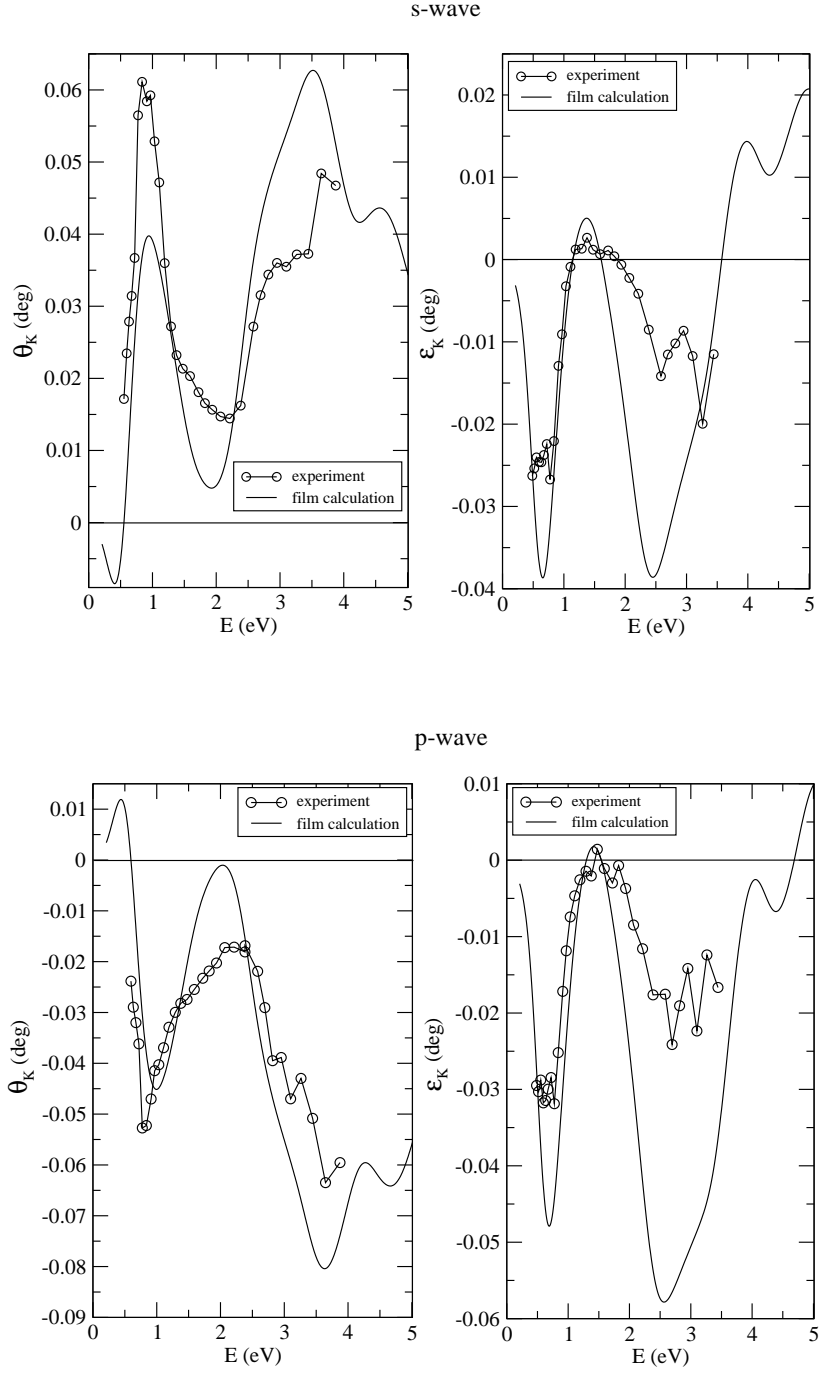

\includegraphics[angle=-0,scale=0.47]{Co2MnSn_swave.eps}\\
\vspace{1.2cm}
\includegraphics[angle=-0,scale=0.47]{Co2MnSn_pwave.eps}
\caption{Theoretical and experimental longitudinal Kerr spectra for Co$_2$MnSn as a function of the photon energy
in both polarizations of the incident radiation.}
\label{fig:tre}
\end{figure}

\begin{figure}
\includegraphics[angle=-90,scale=0.55]{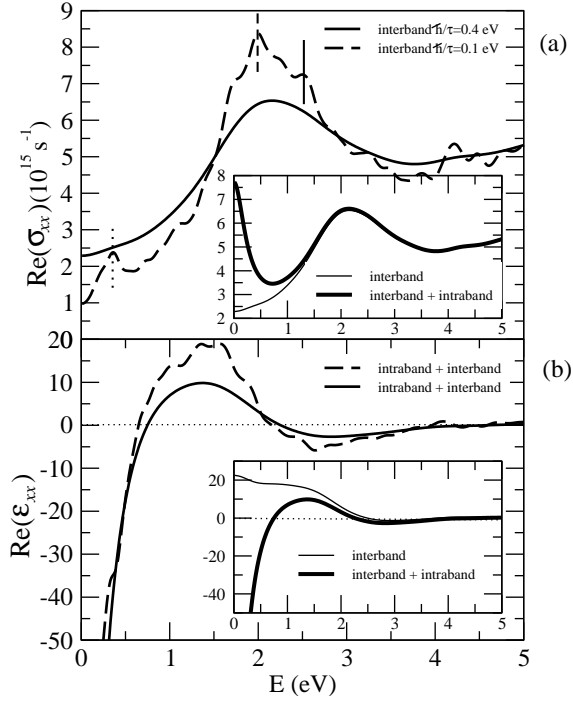}\\
\vspace{1.64cm}
\includegraphics[angle=-90,scale=0.55]{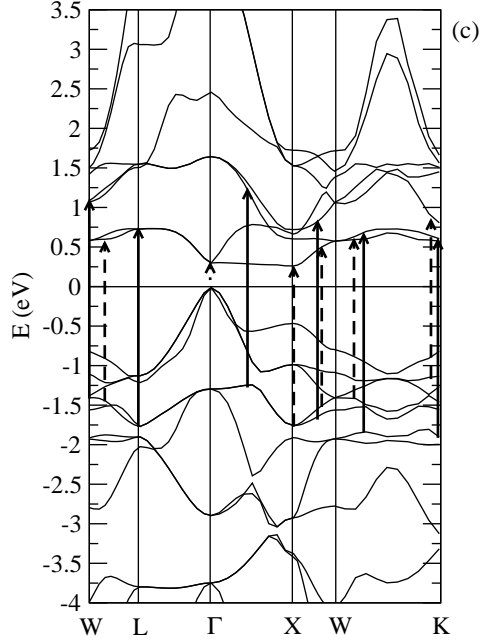}
\caption{Optical properties of \cmg: 
(a) real parts of optical conductivity and (b) dielectric function for different broadenings:
$\frac{\hbar}{\tau}=$ 0.1 eV (dashed) and $\frac{\hbar}{\tau}=$ 0.4 eV (solid). The insets in (a) and (b) show the effect of the
Drude contribution ($\frac{\hbar}{\tau}=$ 0.4 eV, $\hbar\omega_p=$ 3.0 eV and $\frac{\hbar}{\tau_D}=$ 0.2 eV). 
(c) Minority band
structure of \cmg: the arrows (dotted, dashed and solid) mark the transitions 
responsible for the main features highlighted in the optical conductivity, panel (a), at 0.4, 2.0 and 2.5 eV, respectively. 
The zero of the energy scale is set to the Fermi level.}
\label{fig:due}
\end{figure}

Let us first focus on \cmg. The real parts of the optical conductivity,
the dielectric function
and the minority band structure are reported in 
Fig. \ref{fig:due} (a), (b) and (c), respectively.
In order to investigate the relation between the features in the optical
properties and the underlying band structure, we show 
two spectra obtained with a Lorentzian
broadening of $\frac{\hbar}{\tau} = $ 0.4 eV (used in Figs. \ref{fig:f4} and \ref{fig:tre} to compare with
the experimental results) and 0.1 eV (see Eq. (\ref{eq:sigma})), respectively.
The higher resolution of the latter spectrum is useful to trace back more clearly the origin of the transitions,
starting
from the band-structure. In the insets of Fig. \ref{fig:due} (a) and (b), we show
a comparison between the total (interband + intraband) and
interband-only contributions to the optical properties. The arrows (dotted, solid and dashed) 
mark some possible vertical
transitions with an energy corresponding to the
main maxima (at 0.4, 2.0 and 2.5 eV, respectively) in the optical conductivity.
We note that the optical conductivity shows:
{\em i}) a different behavior for energies $<$ 1.5 eV,
depending on whether the Drude contribution is included or not; {\em ii}) 
three main features in the interband spectra (more evident with the smaller
applied broadening), at $\sim$ 0.4, $\sim$ 2.0 and $\sim$ 2.5 eV.
The former consideration is in line with what is generally observed: 
for typical plasma frequencies of a few eV, the
intraband Drude contribution is relevant at small frequencies and becomes negligible
at energies larger than $\sim$ 1.5 eV (given the frequency dependence of Eq. (\ref{eq:drude})). Similarly, the dielectric function
also shows a typical Drude-like behavior at small energies.
In order to better explain the origin of the maxima just
mentioned in the real part of the optical conductivity, 
$\sigma_{xx}$, let us focus on
the minority band-structure of \cmg. 

To better understand the hybridization process, we take a look at the minority spin only:
the interaction between $d$ orbitals of the two Co atoms in the primitive cell forms bonding (antibonding)
states between a 3-fold degenerate $t_{2g}$ ($t_{1u}$) and 2-fold degenerate
$e_g$ ($e_u$) states.
The hybridization between $t_{2g}$ and $e_g$ Co-Co orbitals and Mn $d$ states gives rise\cite{galanak} to bonding and antibonding levels between
3-fold $t_{2g}$ and 2-fold $e_g$, while the $t_{1u}$
and $e_u$ levels do not hybridize with Mn states due to symmetry rules. Therefore, the latter levels are localized at Co sites only.
The hybrid level positions can be inferred from the atom projected density of states (not shown); their energies are:
\begin{equation}\label{eq:level}
e^{\downarrow}_g < t^{\downarrow}_{2g} < t^{\downarrow}_{1u} < E_F < e^{\downarrow}_{u} < e^{\downarrow}_{g} <
t^{\downarrow}_{2g} \ ,
\end{equation}
with a total number of occupied minority spin levels equal to 8 electrons per cell.
In a more ``crude'' picture, $\sigma$ can be obtained simply from the product of the dipole matrix element (taken {\bf k}-independent) and
the joined density of states (JDOS)\cite{ebert} related to the gradient of the energy of the initial and final states.
The highlighted transitions (Fig. \ref{fig:due} panel (c)) at 2.0 and 2.5 eV are relative to excitations from
$t^{\downarrow}_{2g}$ to
$e^{\downarrow}_{u}$ and $e^{\downarrow}_g$ levels around the K and W points (and, to a lower extent, around the X and L
points as well) of the Brillouin zone. Transitions around 0.4 eV and 1.0 eV occur close to the $\Gamma$
point and the $\Gamma$X line exclusively, and involve
electronic excitations from $t^{\downarrow}_{1u}$ to $e^{\downarrow}_{u}$ states.  
The involved bands show a predominantly $d$-character and are pretty flat; thus, we expect a larger contribution coming from
the JDOS rather than the dipole matrix element.

Finally, since we do not have experimental optical data to compare
with, we recall that for the similar
compound Ni$_2$MnSn\cite{kirillova}, a good agreement between theory and experiments
was obtained. More recently, a paper on Ni$_2$MnIn\cite{kudry} showed that the agreement between first-principles
simulations  and optical measurements was sufficiently good, as far as the main
features are concerned, and could be optimized (especially for the peak
energy positions), including some empirical corrections
(e.g., the $\lambda$-fitting\cite{harmon} process) 
 to account for
many-body and excited state effects. Here, we prefer to not consider
any empirical corrections: in fact,
since the agreement with the by far more delicate magneto-optical properties
is already satisfying, we expect that the inclusion
of excited-state corrections would only marginally affect the overall picture which is seen to be
accurate enough already within DFT.

\begin{figure}
\phantom{b}
\vspace{3cm}
\includegraphics[scale=1.2]{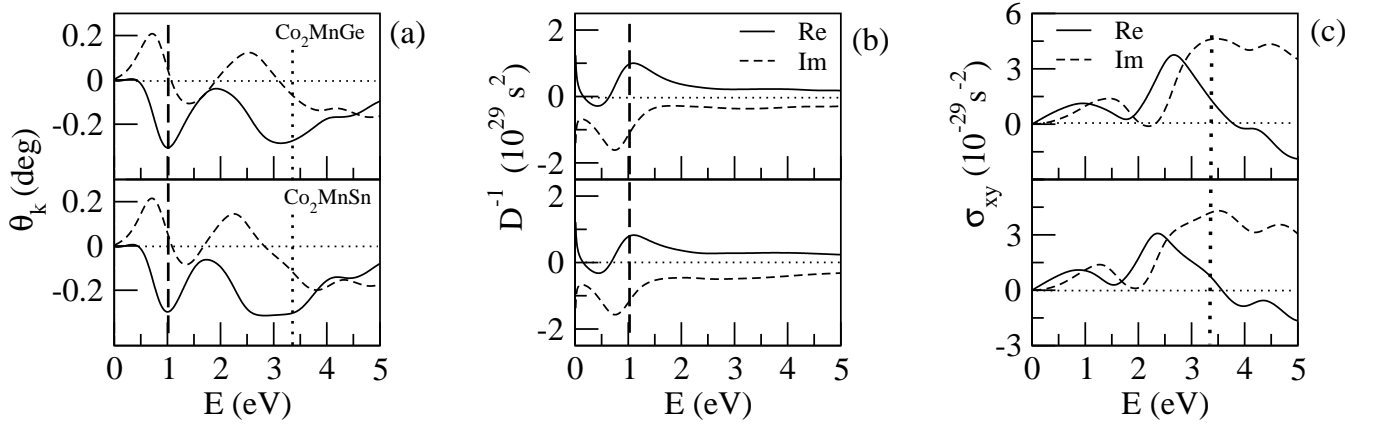}
\caption{(a) Polar Kerr rotation (solid curve) and ellipticity (dashed curve) at normal incidence in Co$_2$MnGe (top panel) and Co$_2$MnSn (bottom panel). 
 (b): inverse denominator in the complex Kerr angle of Eq. (\ref{eq:teta3}).
 (c): off--diagonal conductivity  
  (see numerator of Eq. (\ref{eq:teta3})). In (b) and (c), solid and dashed curves denote real and imaginary parts, respectively.}
\label{fig:uno}
\end{figure}

Having understood the optical properties, we now move to the magneto-optical
properties. Figure \ref{fig:uno} (a) shows the calculated polar spectra (both rotation and ellipticity)
at normal incidence for bulk Co$_2$MnGe and  \cmsn: this is the geometry which provides
the highest Kerr response and,
therefore, it is the one considered to evaluate the performance of a possible magneto-optical application.
The main features of the spectra are similar to those shown in Figs. \ref{fig:f4} and \ref{fig:tre}, relative
to the longitudinal Kerr effect at 45$^{o}$, except for a much larger peak size. 
First of all, the longitudinal Kerr rotation is around 0.05$^{o}$
 (although it should be recalled that the maximum Kerr rotation strictly
depends on the smearing value considered): this is a typical value
for 3$d$ based materials, which are generally not very active from the
magneto-optical point of view.
 Upon reduction of the broadening
values within a physically meaningful range ({\em i.e.} $\frac{\hbar}{\tau} =$ 0.1 eV), the Kerr angle
increases up to 0.07$^{o}$. Still, given the 3$d$ nature of the compounds, 
these are not unexpected results, at variance with
the case of PtMnSb where a giant Kerr rotation was observed both experimentally
\cite{ptmnsb} and theoretically\cite{revantonov}. This is consistent with the
picture given previously\cite{revantonov,ravin} for half-Heusler compounds: in
order to have an extremely high Kerr angle, a large spin-orbit coupling is required in addition to a
large magnetic moment, in this case provided by Mn atoms.

In our samples,  cobalt (a light transition metal) acts as a ``cation",
providing a spin--orbit coupling much smaller than other heavier elements (such as Pt or Au).
On the other hand, also the $p$
element (Ge or Sn) behaves differently from Sb, but the effect of an increased spin-orbit coupling
with increased atomic number of the $p$-electron source element is negligible\cite{slvdefects}, since most of the
states involved in the transitions are localized on the cation sites (Co, Mn).
Our generally small Kerr angles indicate that half-metallicity alone (such as that
typical of \cmsn and \cmg) is not sufficient to guarantee strong MO effects.  
Along the same line, the maximum magnitude of the Kerr angle in Co$_2$MnSn, which deviates from half metallicity,
is not remarkably smaller, but rather of the
same order of magnitude as in \cmg.
Moreover, irrespective of the actual chemical composition,
the Kerr spectra look pretty similar for the two compounds: the Kerr rotation is
generally negative and shows two minima, marked with vertical lines in Fig. \ref{fig:uno}, whereas the Kerr ellipticity exhibits
some zero crossings and two main positive peaks. This has to be related,
of course, to the underlying electronic structure which, as noted before, is very 
similar in the two compounds. 

As shown in Eq. (\ref{boh}), the Kerr angle is given by a strictly magneto-optical numerator,
which depends only on the off-diagonal part of the $\sigma$ matrix (and, therefore, is related to the simultaneous 
presence of exchange and spin-orbit coupling), and by an optical denominator (related
to the diagonal part of the conductivity tensor only). Therefore, some hints
about the origin of the features in the Kerr angles can be gained by plotting,
as reported in Fig. \ref{fig:uno},
the Kerr rotation (panel (a)) and, separately, the numerator (real and imaginary parts, panel (b)) 
and inverse denominator
(real and imaginary parts, panel (c)) of Eq. (\ref{boh}).
In agreement with what was suggested for the half-Heusler compounds
 \cite{revantonov} and also confirmed for a large class of full-Heusler
alloys, we find that
the first minimum (photon energy $\sim$ 1.0 eV) is mostly due to optical contributions (compare with panel (b)),
whereas the second minimum ($\sim$ 3.4 eV) shows a magneto-optical origin (compare with panel (c)).

Finally, we recall that the imaginary part of the off-diagonal conductivity (Fig. \ref{fig:uno}(c))
is directly related to the difference in absorption of the left- (LCP) and
right-circularly polarized (RCP) light. Therefore, the wavelengths characterized by a zero rotation angle
(though slightly dependent on the broadening used) mark the points where 
LCP and RCP-induced transitions have the same overall probability. For both
Co$_2$MnGe and Co$_2$MnSn, this occurs around 2.0-2.3 eV.

\section{Conclusions}
\label{concl}

In summary, we have presented a full characterization of MOKE spectra in \cmg $ \ $  and \cmsn
Heusler
compounds, both from first-principles and experimental points of view.
We have shown that, for a correct comparison between theory and experiment,
the exact measurement conditions have to be taken into account, namely:
Kerr geometry, angle of incidence of the light and multilayered structure
of the samples. This has been achieved by implementing \emph{ab-initio} simulations
in DFT of the tensor $\sigma$ within the Zak model which imposes the correct boundary conditions
in any given geometry and multilayered configuration.
The maximum Kerr rotation results to be of the order of $\sim$0.25$^{o}$, which is
not very appealing for magneto-opto-electronic applications. On the other hand,   
the agreement between theory and experiment allowed a
deep understanding of
the main features in the optical and magneto-optical spectra of these compounds
in terms of the underlying band-structure. This could be very useful in the design of
ad-hoc materials, since a solid basis for predicting their properties and their possible applications
is established. 

\begin{acknowledgements}
Work at Northwestern University supported by the U.S.N.S.F. through its MRSEC Program at the N. U. materials Research Center.
\end{acknowledgements}

\newpage

\begin{thebibliography}{100}
\bibitem{ggavslda} The lattice constants were calculated within the
Generalized Gradient Approximation according to Perdew-Becke-Erzenhof
parametrization, which is known to give lattice constants
close to experiments.
\end{thebibliography}
\begin{table}
\begin{tabular}{|c|c|c|c|c|c|c|}  \hline \hline
& $a$ & $\mu_{tot} $ & $\hbar\omega_p^{\uparrow}$ & $\hbar\omega_p^{\downarrow}$ &
$D^{\uparrow}(E_F)$ & $D^{\downarrow}(E_F)$\\
& (\AA) & $(\mu_B)$ & (eV) & (eV) & (states/eV) & (states/eV)\\ 
\hline \hline
Co$_2$MnGe & 5.74 & 5.00 & 4.4 & 0.0 & 1.5 & 0.0\\ \hline
Co$_2$MnSn & 6.00 & 5.03 & 4.4 & 0.7 & 0.9 & 0.1 \\ \hline
\end{tabular}
\caption{Relevant calculated properties of  Co$_2$MnGe and \cmsn
Heusler compounds: equilibrium lattice constant, total magnetic moment per cell ($\mu_{tot}$), 
 plasma frequency
for up and down spin channels ($\omega_p^{\uparrow}$ and
$\omega_p^{\downarrow}$), and their density of states
at $E_F$ ($D(E_F)^{\uparrow}$ and
$D(E_F)^{\downarrow}$).}
\label{ricap}
\end{table}

%

%
%



\end{document}